%% file: csrcolor.tex
\documentclass[times,doublespace]{cpeauth}
\usepackage{moreverb}
\def\volumeyear{2016}
\usepackage[normalem]{ulem}
\usepackage[capitalize]{cleveref}
\usepackage{amsmath,bm}
\usepackage{amssymb}
\usepackage{graphicx}
\usepackage{listings}
\usepackage{xcolor}
\usepackage{color}
\usepackage{cite}
\usepackage{courier}
\usepackage{float}
\usepackage{pifont}
\usepackage{pbox}
\usepackage{tabu}
\usepackage{xspace}
\usepackage{listings}
\usepackage{algorithm} 
\usepackage{algpseudocode}

\begin{document}
\runningheads{X.~Chen, P.~Li, J.~Fang, T.~Tang, Z.~Wang, C.~Yang}{Efficient and High-quality Sparse Graph Coloring on GPUs}
\title{Efficient and High-quality Sparse Graph Coloring on GPUs}
\author{Xuhao Chen, Pingfan Li, Jianbin Fang, Tao Tang, Zhiying Wang, Canqun Yang}
\address{College of Computer, National University of Defense Technology, Changsha 410073, China}
\corraddr{cxh@illinois.edu}

\input{sections/abstract}
\maketitle
\footnotetext[2]{The source code of this work can be found at
{\texttt{https://github.com/chenxuhao/csrcolor}}}

\input{sections/sect-1}
\input{sections/sect-2}
\input{sections/sect-3}

\input{sections/sect-4}
\input{sections/sect-5}
\input{sections/sect-6}
\bibliographystyle{IEEEtran}
\bibliography{references}
\end{document}

%% file: sections/abstract.tex
\begin{abstract}
Graph coloring has been broadly used to discover concurrency in 
parallel computing. To speedup graph coloring for large-scale datasets, 
parallel algorithms have been proposed to leverage modern GPUs. 
Existing GPU implementations either have limited performance or 
yield unsatisfactory coloring quality (too many colors assigned). 
We present a work-efficient parallel graph coloring implementation
on GPUs with good coloring quality. Our approach employs the \textit{speculative 
greedy} scheme which inherently yields better quality than the method of
finding \textit{maximal independent set}. In order to achieve high 
performance on GPUs, we refine the algorithm to leverage efficient
operators and alleviate conflicts. We also incorporate common 
optimization techniques to further improve performance. Our method 
is evaluated with both synthetic and real-world sparse graphs on the 
NVIDIA GPU. Experimental results show that our proposed implementation 
achieves averaged 4.1$\times$ (up to 8.9$\times$) speedup over the 
serial implementation. It also outperforms the existing GPU implementation 
from the NVIDIA CUSPARSE library (2.2$\times$ average speedup), 
while yielding much better coloring quality than CUSPARSE. 
\end{abstract}


%% file: sections/sect-1.tex
\section{Introduction}
Graph processing algorithms are getting a growing research interest
in the past decade. They are pervasively used in many application 
domains, such as scientific computing, social networks, simulations 
and bioinformatics. Parallelizing graph algorithms is challenging
because of their inherent irregularity. To leverage modern
massively parallel processors, e.g. GPUs, makes the problem even 
harder because of the difficulty of managing massive hardware resources
and sophisticated memory hierarchies. In this paper, we investigate
the problem of graph coloring which assigns colors to all the vertices
of a graph such that no neighboring vertices have the same color.
Graph coloring is a fundamental graph algorithm that has been 
employed in many applications~\cite{TT,LSS,EDD,Chaitin,Berchtold}, 
and is also intensively utilized by scientific computing 
to discover concurrency, e.g. high performance conjugate gradient 
(HPCG)~\cite{HPCG} and incomplete-LU factorization~\cite{ILU}, 
where coloring is used to identify subtasks that can be carried out 
or data elements that can be updated simultaneously.

To deal with large-scale datasets, parallel graph coloring
algorithms~\cite{compare,PGC} have been proposed to leverage 
the massive hardware resources on modern multicore CPUs or GPUs. 
Existing parallel implementations of graph coloring can be classified 
into two categories: 1) speculative greedy (SGR) scheme based~\cite{GCA} 
and 2) maximal independent set (MIS) based~\cite{MIS}. There are existing 
GPU implementations of both catigories. With different algorithms, they 
exhibit different characteristics of performance and coloring quality.
MIS implementations~\cite{ILU} are usually fast since multiple threads 
can find MIS in parallel independently, and more importantly they 
can substantially reduce the total number of memory accesses. 
But they inherently yield too many colors. On the other hand, 
SGR implementations~\cite{EGC} generally use fewer colors than 
MIS ones, but without careful mapping and 
optimizations, they spend much more time to complete coloring. 

To overcome the limitations of existing approaches, we propose a high 
performance GPU graph coloring implementation which can produce 
high-quality coloring. Our method is built based on the SGR scheme 
so that good coloring quality is guaranteed. It is then optimized 
specifically for the GPU architecture to improve performance. We choose
data-driven instead of topology-driven mapping strategy for better work 
efficincy, and make algorithm tradeoffs to leverage efficient operators
and alleviate the side effects of massive parallelism on GPUs. Meanwhile, 
we incorporate common optimization techniques, e.g. kernel fusion,
to further improve performance. The major insight of 
this work is that \textit{algorithm-specific optimizations} are as 
important as common optimization techniques for high performance 
graph algorithm on GPUs. The main contributions of this paper are:

1) We present a work-efficient GPU graph coloring algorithm based 
on the speculative greedy scheme. The algorithm is carefully refined
to better leverage GPU's bulk-synchronous model. It shows the importance
of algorithm refinement to achieve high performance on GPUs.

2) We employ optimization techniques specifically for the GPU architecture 
to take advantage of GPU's computation resources and memory hierarchies.
Our practice further demonstrates GPU's capability on accelerating graph algorithms.

The rest of the paper is organized as follows:
the existing serial and parallel algorithms as well as the state-of-the-art
GPU implementations are introduced in Section~\ref{sect:motivation}.
Our proposed design is presented in Section~\ref{sect:design}.
We present the experimental results in Section~\ref{sect:evaluation}.
Section~\ref{sect:relatedwork} discusses related work,
and Section~\ref{sect:conclusion} concludes.

%% file: sections/sect-2.tex
\section{Background and Motivation}\label{sect:motivation}
The graph coloring problem refers to the assignment of colors to elements 
(vertices or edges) of a graph subject to certain constraints. In this paper, 
we focus on \textit{vertex coloring} which assigns colors to vertices so 
that no two neighboring (connected) vertices are 
assigned the same color. There are several known applications of graph 
coloring, such as time-tabling and scheduling~\cite{TT,LSS,EDD}, 
register allocation~\cite{Chaitin}, high-dimensional nearest-neighbor 
search~\cite{Berchtold}, sparse-matrix computation~\cite{ILU,HPCG} 
and assigning frequencies to wireless access points~\cite{FA}.

Graph coloring that minimizes the number of colors is a 
NP-complete problem, and is known to be NP-hard even solved 
approximately~\cite{LDE}. In this paper, we focus on 
\textit{approximate graph coloring} which yields near-optimal coloring 
quality. Many heuristics have been developed for approximate solutions, 
including First Fit (FF), Largest Degree First (LF), etc. These heuristics 
make trade-offs between minimizing the number of colors
and execution time but generally faster algorithms have poor 
coloring quality while slower ones tend to yield fewer colors. In the 
following, we introduce some existing sequential and parallel algorithms.

\subsection{Sequential Graph Coloring}
A sequential algorithm~\cite{GCA,CUSP} based on the greedy scheme 
is shown in Algorithm~\ref{alg:sequential}. In all the algorithms 
specified in this paper, we use similar data structures to those 
introduced in \cite{GCA}. $adj(v)$ denotes the set of vertices adjacent 
to the vertex $v$, $color$ is a vertex-indexed array that stores the color of 
each vertex, and $colorMask$ is a color-indexed mask array used to mark the 
colors that are impermissible to a particular vertex $v$. At the beginning 
of the procedure, the array $color$ is initialized with each entry $color[w]$ 
set to zero to indicate that vertex $w$ is not yet colored, and each entry 
of the array $colorMask$ is initialized with some value $a \notin V$.
When processing the vertex $v$, the algorithm scans all its neighbors (line 3),
and their colors are forbidden to be assigned to the vertex $v$ (line 4). 
By the end of the inner for-loop, all of the colors that are 
impermissible to the vertex $v$ are recorded in the array $colorMask$. 
It is then scanned from left to right to search the lowest 
positive index $i$ at which a value different from the current 
vertex $v$ is encountered; this index corresponds to the 
smallest permissible color $c$ to the vertex $v$ (line 6). 
The color $c$ is then assigned to the vertex $v$ (line 7). 

\begin{algorithm}[!t]
\caption{Sequential Greedy Algorithm~\cite{GCA}}
\label{alg:sequential}
\begin{algorithmic}[1]
\Procedure{Greedy($G(V,E)$)} {}
\For{each vertex $v \in V$}
	\For{each vertex $w \in adj(v)$}
		\State $colorMask[color[w]] \leftarrow v$
	\EndFor
	\State $c \leftarrow \min{\{i > 0: colorMask[i] \neq v\}}$
	\State $color[v] \leftarrow c$
\EndFor
\EndProcedure
\end{algorithmic}
\end{algorithm}

\subsection{Parallel Graph Coloring}
Parallel graph coloring has been applied to large-scale problems, 
such as sparse-matrix computation~\cite{HPCG,ILU} and chromatic 
scheduling~\cite{EDD} to meet the performance requirement. 
Because of its sequential nature, the greedy scheme is 
challenging to parallelize. Basically, two classes of approaches 
have been proposed in the past to tackle this issue. 

\begin{algorithm}[!t]
\caption{Parallel GM Algorithm~\cite{GCA}}
\label{alg:gm}
\begin{algorithmic}[1]
\Procedure{GM($G(V,E)$)} {}
\State $W \leftarrow V$ \Comment{Initialize the worklist}
\While {$W \neq \varnothing$}
  \For{each vertex $v \in W$ in \texttt{parallel}}
	  \For{each vertex $w \in adj(v)$}
		  \State $colorMask[color[w]] \leftarrow v$
  	\EndFor
  	\State $c \leftarrow \min{\{i > 0: colorMask[i] \neq v\}}$
  	\State $color[v] \leftarrow c$
  \EndFor
	\State $R \leftarrow \varnothing$ \Comment{Initialize the remaining worklist}
	\For{each vertex $v \in V$ in \texttt{parallel}}
	  \For{each vertex $w \in adj(v)$}
		  \If{$color[v] = color[w]$ and $v < w$} 
				\State $R \leftarrow R \cup \{v\}$ 
			\EndIf
		\EndFor
	\EndFor
  \State $W \leftarrow R$ \Comment{Update the worklist}
\EndWhile
\EndProcedure
\end{algorithmic}
\end{algorithm}

Gebremedhin and Manne (GM)~\cite{PGC} used \textit{speculation} 
to deal with the inherent sequentiality of the greedy scheme. 
It colors as many vertices as possible in parallel, tentatively 
tolerating potential conflicts, and resolve conflicts afterwards. 
Algorithm~\ref{alg:gm} shows the details of the GM algorithm. 
It can be divided into two parts: the first part (from line 4 
to line 10) is the same as the sequential algorithm but done in parallel. 
The second part (from line 12 to line 18) does the conflict resolve
(line 14) and puts the conflicting vertices into the remaining worklist
(line 15). Based on this \textit{speculative greedy} (SGR) algorithm, 
\c{C}ataly\"{u}rek~\emph{et~al.} developed OpenMP implementations for 
the multi-core and massively multithreaded architectures~\cite{GCA}. 
Rokos~\emph{et~al.} improved \c{C}ataly\"{u}rek's algorithm and 
implemented it on the Intel{\textregistered} Xeon Phi coprocessor~\cite{Rokos}.

The other approach relies on iteratively finding a \textit{maximal independent set}
(MIS) of vertices in a progressively shrinking graph and coloring the vertices 
in the independent set in parallel. In many of the methods in this class, 
the independent set is computed in parallel using some variant of Luby's 
algorithm~\cite{MIS}. An example is the work of Jones and Plassmann 
(JP)~\cite{JPL}. Algorithm~\ref{alg:jp} shows the details of the JP algorithm. 
Gjertsen~\emph{et~al.}~\cite{PHI} introduced an advanced parallel heuristic, 
PLF, that consistently generates better colorings than the JP heuristic with 
slight overhead. Two new parallel color-balancing heuristics, PDR(k) and 
PLF(k) are also introduced. Hasenplaugh~\emph{et~al.}~\cite{OHP} 
futher improve the ordering heuristics based on the JP algorithm.

\subsection{CUDA programming and GPU Graph Coloring}
With the success of CUDA~\cite{CUDA} programming model, general-purpose 
graphics processing units (GPGPUs)~\cite{GPGPU} have been widely used for 
high performance computing (HPC) and many other application domains during 
the last decade. 
In CUDA, individual functions 
executed on the GPU device are called {\it kernel} functions, written in a single 
program multiple-data (SPMD) form. Each instance of the SPMD function is executed 
by a GPU {\it thread}. Groups of such threads, called {\it thread blocks}, are 
guaranteed to execute concurrently on the same streaming multiprocessors (SMs).
Within each group, subgroups of threads called {\it warps} are executed in 
lockstep, evaluating one instruction for all threads in the warp at once.
One of the major difficulties of CUDA programming is to manage the GPU memory
hierarchy. It consists of register files, L1 memories (scratchpad, L1 cache, 
and read-only data cache), the shared L2 cache, and the off-chip GDDR DRAM~\cite{Kepler}. 
Scratchpad memory (\textit{shared memory} in CUDA terminology) is programmer 
visible and can be used for explicit intra thread block communication.
L2 cache works as the central point of coherency, and is shared 
across all threads of the entire kernel. 

\setlength{\textfloatsep}{10pt}
\begin{algorithm}[!t]
	\caption{Parallel JP Algorithm~\cite{ILU}}
	\label{alg:jp}
	\begin{algorithmic}[1]
		\Procedure{JP($G(V,E)$)} {}
		\State $W \leftarrow V, c \leftarrow 1$
		\While {$W \neq \varnothing$}
		\State $S \leftarrow \varnothing$ \Comment{Initialize the independent set}
		\For{each vertex $v \in W$ in \texttt{parallel}}
		\State $r(v) \leftarrow random()$ 
		\EndFor
		\For{each vertex $v \in W$ in \texttt{parallel}}
		\State $flag \leftarrow true$
		\For{each vertex $w \in adj(v)$}
		\If{$r(v) <= r(w)$}
		\State $flag \leftarrow false$
		\EndIf
		\EndFor
		\If{$flag = true$}
		\State $S \leftarrow S \cup \{v\}$
		\EndIf
		\EndFor
		\For{each vertex $v \in S$ in \texttt{parallel}}
		\State $color[v] \leftarrow c$ \Comment{Color an independent set}
		\EndFor
		\State $W \leftarrow W - S, c \leftarrow c + 1$
		\EndWhile
		\EndProcedure
	\end{algorithmic}
\end{algorithm}

Several GPU graph coloring implementations have been proposed so far using 
either GM or JP algorithm. Grosset~\emph{et~al.}~\cite{EGC} implement the GM 
algorithm using CUDA. They use a 3-step graph coloring framework: 
1) \textit{Graph partitioning} which partitions the graph into subgraphs and 
identifies boundary vertices, 2) \textit{graph coloring \& conflicts detection} 
which colors the graph using the specified heuristic, e.g. FF, and identifies 
color conflicts, and 3) \textit{sequential conflicts resolution} which goes 
back to CPU and resolves the conflicts. Note that step 2 is performed multiple 
times on GPU to reduce the number of conflicts before going back to CPU. 
Although this \texttt{3-step GM} algorithm assigns as few colors as 
the serial algorithm, its performance is poor, or even 
worse than the sequential graph coloring for many datasets, 
meaning the GPU computation horsepower is not leveraged very well.

The CUSPARSE~\cite{CUSPARSE} library offered by NVIDIA includes 
a \texttt{csrcolor}~\cite{ILU} routine which does graph coloring 
on a given graph in CSR format~\cite{CSR}. The algorithm of 
\texttt{csrcolor} is derived from the JP algorithm, but uses the 
\textit{multi-hash} method to find independent sets. Basically, 
several hash functions (instead of random number generators)
are selected, and used to generate hash values for each vertex 
with the vertex number as the input of the hash functions. 
Given the generated hash values, local maximum and minimum values 
can be found, and distinct (maximal) independent sets are generated 
for each of the hash values. Assume $N$ hash values are associated 
with each vertex, and used to create different pairs of (maximal) 
independent sets, this multi-hash method can generate $2N$ (maximal) 
independent sets at once. Compared to the GM algorithm, this method
significantly reduces accesses to the $color$ array, because it
compares the generated hash values (in the registers) instead of the 
colors of neighbors (in the memory). As reported~\cite{ILU}, the \texttt{csrcolor} 
implementation runs pretty fast on modern NVIDIA GPUs. However, it usually 
produces several times more colors than the sequential algorithm, which is 
not satisfactory for many applications. For example, when applied to exploiting 
concurrency in parallel computing, more colors means less parallelism,
because tasks (vertices) with the same color can be processed concurrently. 

\begin{figure}[!t]
	\begin{center}
		\includegraphics[width=0.8\textwidth]{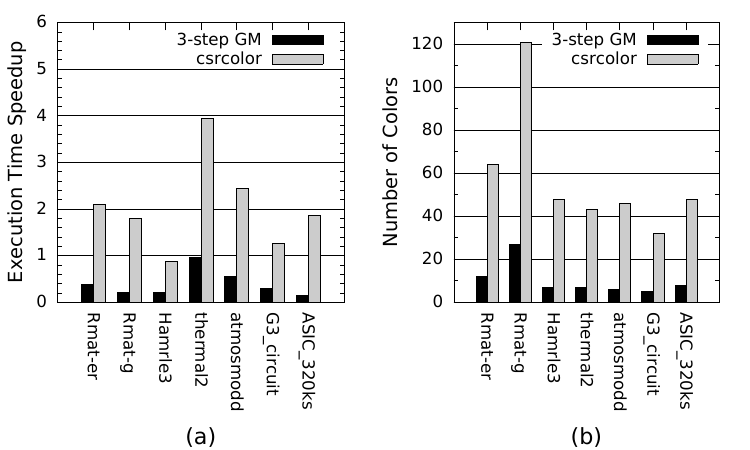}
		\caption{Comparison between two existing GPU graph coloring implementations: 
			\texttt{3-step GM} and \texttt{csrcolor}. (a) performance, i.e. runtime 
			speedup normalized to the serial implementation (the more the better); 
			(b) coloring quality, i.e. the number of colors assigned (the less the better). 
			This figure shows that existing GPU implementations either have poor 
			performance or yield unsatisfactory coloring quality, which motivates our work.}
		\label{fig:motivation}
	\end{center}
\end{figure}

We evaluate the two existing GPU implementations of graph coloring
on the NVIDIA K40c GPU. \cref{fig:motivation} shows the performance 
and coloring quality of both implementations. As illustrated, 
\texttt{3-step GM} yields much better coloring quality than \texttt{csrcolor}, 
but its performance is even worse than the sequential implementation, 
meaning it does not exploit GPU hardware very well. On the other hand, 
\texttt{csrcolor} runs much faster than \texttt{3-step GM}, and gains a 
certain degree of speedup over the sequential implementation. However, 
this good performance comes at the expense of much worse coloring quality: 
it yields several times more colors than the sequential implementation and 
\texttt{3-step GM}. The limitations of \texttt{csrcolor} and \texttt{3-step GM} 
motivate us to design a better implemention of parallel graph coloring 
for GPUs to achieve both high performance and good coloring quality.

%% file: sections/sect-3.tex
\section{Design}\label{sect:design}
Graph algorithms are typical irregular algorithms~\cite{irregular} 
that are considered to be difficult to parallelize on GPUs. However, 
recent works~\cite{Merrill,BC,SCC,Morph,MST,SSSP} show that 
GPUs are capable to substantially accelerate graph algorithms if 
they are carefully designed and optimized for the GPU architecture.
Although the previously proposed optimization techniques for other
graph algorithms can be applied to graph coloring, we show that
refining the algorithm for GPUs is essential for our case.

\setlength{\textfloatsep}{10pt}
\algdef{SE}[DOWHILE]{Do}{doWhile}{\algorithmicdo}[1]{\algorithmicwhile\ #1}

\begin{algorithm}[!t]
	\caption{FirstFit routine}\label{alg:ff}
	\begin{algorithmic}[1]
		\Function{FirstFit} {$v$}
		\For{each vertex $w \in adj(v)$}
		\State $colorMask[color[w]] \leftarrow v$
		\EndFor
		\State $c \leftarrow \min{\{i > 0: colorMask[i] \neq v\}}$
		\State $color[v] \leftarrow c$
		\EndFunction
	\end{algorithmic}
\end{algorithm}

\begin{algorithm}[!t]
\caption{ConflictResolve routine}\label{alg:cr}
\begin{algorithmic}[1]
\Function{ConflictResolve} {$v$}
	\For{each vertex $w \in adj(v)$}
	\If{$color[v]$ = $color[w]$ and $v < w$}
		\State $color[v] \leftarrow 0$
	\EndIf
	\EndFor
\EndFunction
\end{algorithmic}
\end{algorithm}

\begin{figure}[!b]
	\begin{center}
		\includegraphics[width=0.7\textwidth]{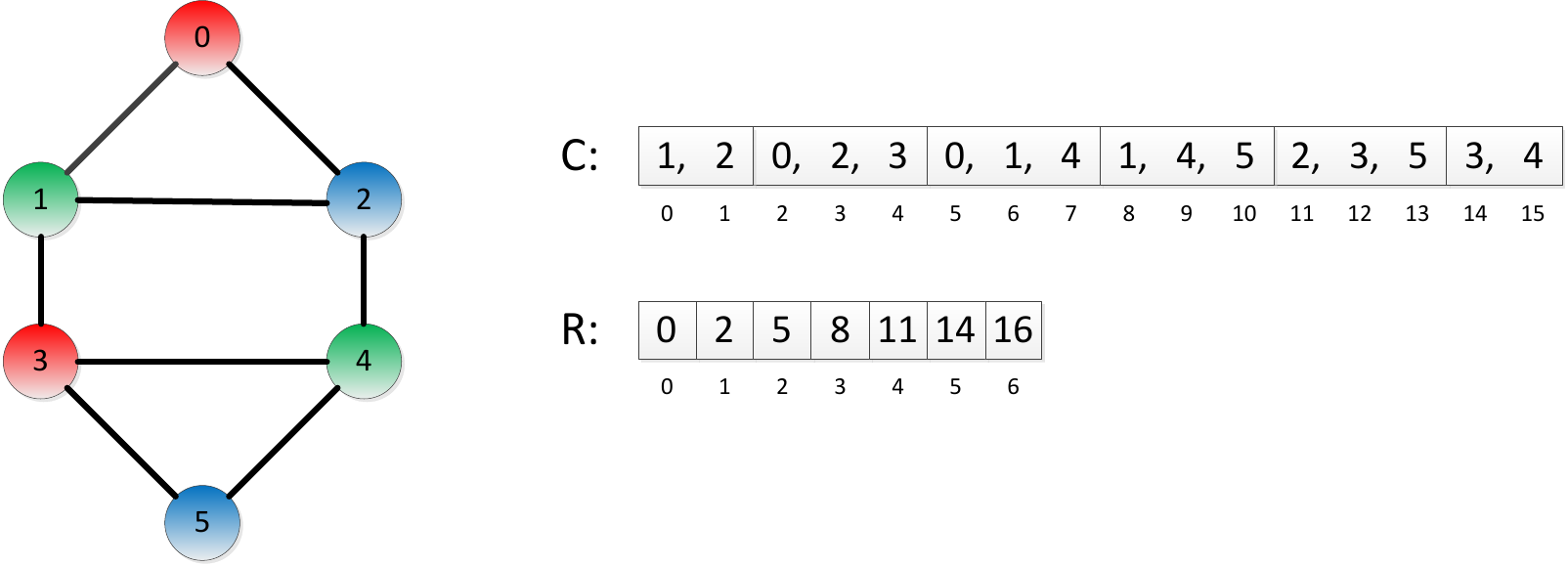}
		\caption{An example of the compressed sparse row (CSR) format.
			For this graph, at least three colors (red, green, blue) are needed.}
		\label{fig:csr}
	\end{center}
\end{figure}

As mentioned in Algorithm~\ref{alg:gm}, the graph coloring workload 
is composed of two major components: assign the first permissible 
color (Algorithm~\ref{alg:ff}. \texttt{FirstFit}) and resolve conflicting 
vertices (Algorithm~\ref{alg:cr}. \texttt{ConflictResolve}). The operations 
are trivial, but GPU's massively parallel model makes it challenging 
to efficiently parallelize these workloads. We investigate the two 
activities in the following analyses using NVIDIA Tesla K40c GPUs.

Note that we use the well-known compressed sparse row (CSR)~\cite{CSR} 
sparse matrix format to store the graph in memory consisting of two arrays. 
\cref{fig:csr} provides a simple example. The column-indices array $C$
is formed from the set of the adjacency lists concatenated into a single 
array of $m$ ($m$ is the number of edges) integers. The row-offsets 
$R$ array contains $n + 1$ ($n$ is the number of vertices) integers, and 
entry $R[i]$ is the index in $C$ of the adjacency list of the vertex $v_{i}$.

\subsection{The Baseline Design}

\begin{algorithm}[!t]
\caption{Topology-driven Parallel Graph Coloring}
\label{alg:topo}
\begin{algorithmic}[1]
\Procedure{Topo-GC($G(V,E)$)} {}
\Do
  \State $changed \leftarrow false$
  \For{each vertex $v \in V$ in \texttt{parallel}}
    \If{$color[v] = 0$} \Comment{Not colored yet}
      \State \Call{FirstFit}{$v$}
      \State $changed \leftarrow true$
    \EndIf
  \EndFor
  \For{each vertex $v \in V$ in \texttt{parallel}}
    \If{$colored[v]$ = false} \Comment{Not Colored yet}
      \State \Call{ConflictResolve}{$v$}
	  \If{$v$ is not confliting}
	    \State $colored[v]$ = true;
	  \EndIf
    \EndIf
  \EndFor
\doWhile {$changed = true$}
\EndProcedure
\end{algorithmic}
\end{algorithm}

In the previous evaluation we find that speculative greedy (i.e. GM) 
algorithm inherently yields better coloring quality than the maximal 
independent set (i.e. JP) method. Thus we choose to use the speculative 
greedy scheme and design our baseline algorithm on top of it.
Compared to the \texttt{3-step GM} algorithm, our proposed
GPU implementation maps the entire coloring work onto the GPU,
consequently removing the data transfer between the CPU and the GPU,
while the CPU is only responsible for controlling the progress. The 
rationale behind this change of mapping is that throughput-oriented 
processors are good at exploiting data-level parallelism and thus
recomputing the conflicting vertices rather than serializing it onto 
the CPU would be more straightforward and efficient. 

\begin{algorithm}[!t]
	\caption{Data-driven Parallel Graph Coloring}
	\label{alg:data}
	\begin{algorithmic}[1]
		\Procedure{Data-GC($G(V,E)$)} {}
		\State $W_{in} \leftarrow V$ \Comment{Initialize the $in$ worklist}
		\While {$W_{in} \neq \varnothing$}
		\For{each vertex $v \in W_{in}$ in \texttt{parallel}}
		\State \Call{FirstFit}{$v$}
		\EndFor
		\State $W_{out} \leftarrow \varnothing$ \Comment{Initialize the $out$ worklist}
		\For{each vertex $v \in W_{in}$ in \texttt{parallel}}
		\State \Call{ConflictResolve}{$v$}
		\If{$v$ is confliting}
		\State $W_{out} \leftarrow W_{out} \cup \{v\}$ \Comment{Atomic push}
		\EndIf
		\EndFor
		\State $swap(W_{in}, W_{out})$ \Comment{Swap the worklists}
		\EndWhile
		\EndProcedure
	\end{algorithmic}
\end{algorithm}

Nasre~\emph{et~al.}~\cite{DVT} introduced the concept of \textit{topology-driven} 
and \textit{data-driven} imlementations of irregular applications on GPUs.
For graph algorithms, the topology-driven implementation simply 
maps each vertex to a thread, and in each iteration, the thread 
stays idle or is responsible to process the vertex depending on
whether the corresponding vertex has been processed or not.
The topology-driven implementation is straightforward, and 
since GPUs are suitable for accelerating data-parallel applications, 
it is easy to map onto the GPU hardware and possibly get speedup.
By contrast, the data-driven implementation maintains a worklist
which holds the remaining vertices to be processed. In each iteration, 
threads are created in proportion to the size of the worklist (i.e. the 
number of vertices in the worklist). Each thread is responsible for
processing a certain amount of vertices in the worklist, and no thread 
is idle. Therefore, the data-driven implementation is generally more
work-efficient than the topology-driven one, but it needs extra overhead
to maintain the worklist. Note that the data-driven implementation
still suffers from load imbalance problem, since vertices may have 
different amount of edges to be processed by the corresponding threads.

We implement graph coloring in these two fashions. Algorithm~\ref{alg:topo} 
shows the topology-driven graph coloring algorithm. In this topology-driven 
algorithm, a flag $changed$ is used to indicate whether all the vertices are 
colored or not. It is cleared at the beginning of each iteration, and set 
by one or more threads if any vertex is colored. Once all the vertices 
have been colored, the flag remains $false$ and the algorithm finally 
terminates. Both \texttt{FirstFit} and \texttt{ConflictResolve} are 
similar to those in the GM algorithm, but in \texttt{ConflictResolve} 
a bitmask $colored$ is used to avoid recomputation.
Algorithm~\ref{alg:data} shows the data-driven graph coloring algorithm. 
It is almost the same as the GM algorithm except that Algorithm~\ref{alg:data} 
uses \textit{double buffering}~\cite{DVT} to avoid copying the worklist. 
The two worklists $W_{in}$ and $W_{out}$ are referenced by pointers, and they 
are swapped at the end of each iteration. Since they are operated using pointers 
instead of data values, no copy operation is required between the two worklists. 

\textbf{Atomic Operation Reduction.} In Algorithm~\ref{alg:data}, since the 
$out$ worklist is a shared data structure, pushing elements into the worklist 
(line 11) requires atomic operations to ensure correctness. Although GPU architects 
have paid a lot of effort to optimize atomic operation, serialization from atomic 
synchronization is still expensive for GPUs~\cite{Merrill}. Merrill~\emph{et~al.}
~\cite{Merrill} proposed to use software \textit{prefix sum}~\cite{Blelloch,Sengupta} 
for updating the shared worklist. Given a list of allocation requirements for 
each thread, prefix sum computes the offsets for where each thread should start 
writing its output elements. Fortunately, efficient GPU prefix sums~\cite{StreamScan} 
have been proposed, and the CUB~\cite{CUB} library has already provided standard 
routines for CUDA users to invoke. Thus we need only one atomic operation for each block.

\textbf{Color Clearing.} In Algorithm~\ref{alg:data}, when a vertex is determined
to be conflicting, it is pushed into the worklist. Intuitively, its color
should be cleared and it will be assigned color in the next iteraion.
However, functionally it is not a necessary operation. In the CPU parallel
algorithm, this is not an issue. But for the GPU implementation, it is important
to clear the color, so that when its neighbors check its color, there won't
be conflicts. Thus in Algorithm~\ref{alg:cr}, the color is cleared (line 4).
We observe non-trivial performance drop if the operation is removed.
In the following sections we will see that the techniques to 
alleviate conflicts are performance critical to our GPU implementations.

\begin{figure}[!t]
	\begin{center}
		\includegraphics[width=0.7\textwidth]{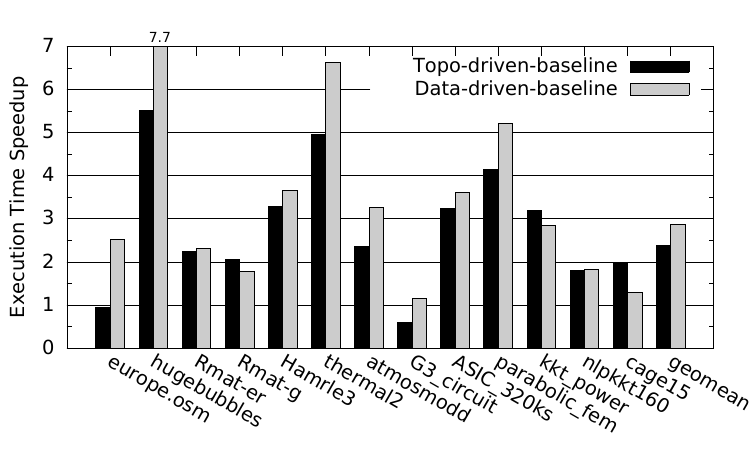}
		\caption{Runtime speedup of topology-driven and data-driven 
			implementations, normalized to the sequential implementation.}
		\label{fig:topo-vs-data}
	\end{center}
\end{figure}

\cref{fig:topo-vs-data} compares their performance. As shown 
in the figure, the data-driven implementation outperforms the 
topology-driven one on average, although the latter is more intuitive to 
implement on the GPU. This is easy to understand because the parallelism 
decreases in graph coloring as the iteration moves forward, and the 
topology-driven implementation has plenty of threads with no work to do, 
while the data-driven implementation is work-efficient although mantaining 
the worklist costs extra overhead. In the following discussion, we 
take this data-driven implementation as our baseline implementation.
To achieve higher performance, we refine the algorithm to 
alleviate the side effects of massive parallelism and leverage 
efficient operators. We call them \textit{algorithm-specific
optimizations}. We also employ common (non-algorithm-specific) 
optimization techniques in Section~\ref{subsect:cot}.

\subsection{Algorithm-specific Optimizations}

As most parallel graph processing algorithms, parallel graph coloring is
iterative. Therefore it is important to ensure quick convergence for high
performance. In the case of graph coloring, the number of iterations 
required to complete coloring highly depends on the conflict situation.
For dense graphs, conflicts happen so frequently that no parallel
algorithm can efficiently solve the problem. Our work thus focuses on 
sparse graph coloring which is more common in real-world applications.
Even so it is still challenging to parallelize it on GPUs, because the
thousands of threads in the massively parallel programming model make
the conflicts happen much more frequently. This is not an issue on CPUs
since there are only several or dozens of threads running simultaneously.
We propose \textit{heuristic conflict resolve} and employ \textit{thread 
coarsening} technique to alleviate this side-effect of GPU parallelism.

\begin{figure}[!t]
	\begin{center}
		\includegraphics[width=0.7\textwidth]{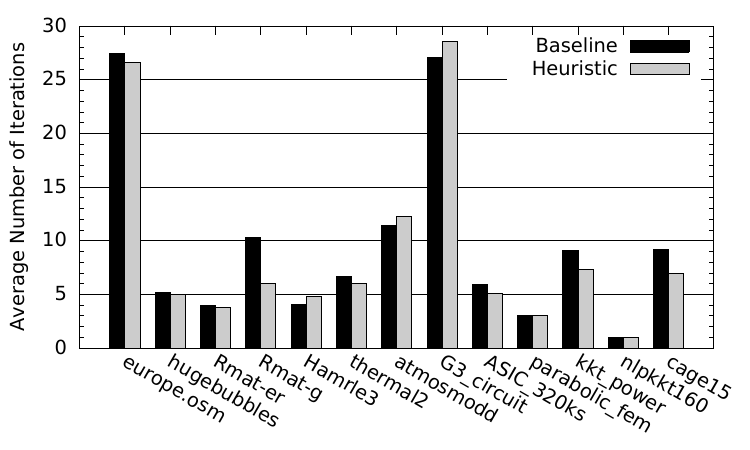}
		\caption{Average number of iterations with the baseline and 
		heuristic implementations. Faster convergence leads to an 
		average of 10.4\% entire program speedup over the baseline.}
		\label{fig:iteration}
	\end{center}
\end{figure}

\textbf{Heuristic Conflict Resolve.}
To reduce the number of iterations, an important part is to reduce conflicts.
Since conflicts happen when two adjacent vertices are assigned the same color,
deciding which of the two conflicting vertices to be re-assigned in the next
iteration affects the following conflict situation. An intuitive scheme
is to pick the one with smaller or larger vertex id, but this is
surely far away from optimal. We apply \textit{heuristic conflict resolve}
that prioritizes coloring the vertex with larger degree and puts the smaller
one into the worklist to be processed in the next iteration. The rational 
behind this heuristic is that vertices with larger degrees have more neighbors
and thus are more likely to cause conflicts in the future. So it is better
to color large-degree vertices first and reduce the possibility of conflicts.
When the two vertices have the same degree, the one with smaller vertex id is picked.
\cref{fig:iteration} illustrates the average number of iterations required
to complete coloring. It is shown that benchmarks e.g. \texttt{rmat-g}
and \texttt{cage15} can have significant iteration reduction using the 
heuristic. We also observe 43\% and 50\% execution time speedup of 
the entire program for the two benchmarks compared to the baseline. 
On average, the heuristic yields 10.3\% speedup over the baseline.

\begin{figure}[!t]
	\begin{center}
		\includegraphics[width=0.7\textwidth]{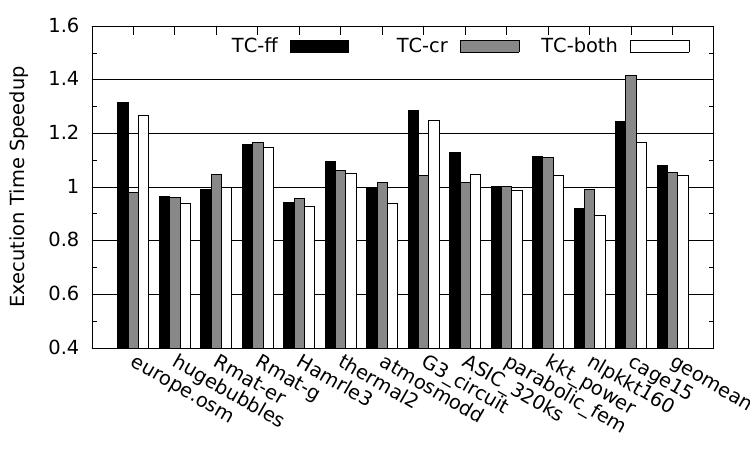}
		\caption{Program execution time speedup of \texttt{thread coarsening}
		on \texttt{FirstFit} (TC-ff), \texttt{ConflictResolve} (TC-cr) and 
		both kernels (TC-both), all normalized to the baseline.}
		\label{fig:tc}
	\end{center}
\end{figure}

\textbf{Thread Coarsening.}
\textit{Thread coarsening} is a common technique utilized in CUDA or OpenCL
programs. It merges several threads together and thus
have each thread do more work. This reduces the total number of threads and
directly affects how data parallel work is mapped to the underlying hardware.
Usually it is used to reduce the amount of redundant computation and thus
can improve performance. For our case, however, it is used to reduce conflicts,
since massive amount of threads on the GPU cause severe conflicts which is
not an issue on the CPU. \cref{fig:tc} shows the effect of thread coarsening 
applied to \texttt{FirstFit}, \texttt{ConflictResolve} or both. Here we launch
$nSM \times max\_blocks$ thread blocks, where $nSM$ is the number of SMs on
the GPU and $max\_blocks$ is the maximum number of thread blocks that is allowed
to be launched on each SM. $max\_blocks$ depends on how many resources (e.g. 
registers, shared memory) a thread block allocates. Each block has 128 threads.
Note that this is not the optimal configuration which is different for different 
benchmarks, and some benchmarks would be faster with even fewer thread blocks.
Autotuning techniques would be helpful, but this is out of the range of this paper.
As shown, benchmarks e.g. \texttt{G3\_circuit} and \texttt{cage15} can 
remarkably benefit from thread coarsening. On average, applying thread coarsening 
on both kernels can improve performance by 4.4\% over the baseline.

\textbf{Bitset Operation.} 
Another major time-consuming part stems from writes and reads on the
$colorMask$ data structure in the \texttt{FirstFit} kernel. For each
vertex, all its neighbors are visited to collect impermissible colors
which are written into $colorMask$. This information is then sequentially
checked to find the first permissible color. In the worst case, all
the elements in the $colorMask$ array are checked, but actually we only
need to find one permissble color. To reduce the costs of this operation, 
we propose to use \texttt{bitset} operations to implement reads and writes 
on the $colorMask$ array. \texttt{bitset} is a standard class template in C++, 
but no similar support is provided in CUDA yet. Thus we implement similar
operations to mimic the functionality of the \texttt{bitset} class. 

\begin{figure}[!t]
	\begin{center}
		\includegraphics[width=0.7\textwidth]{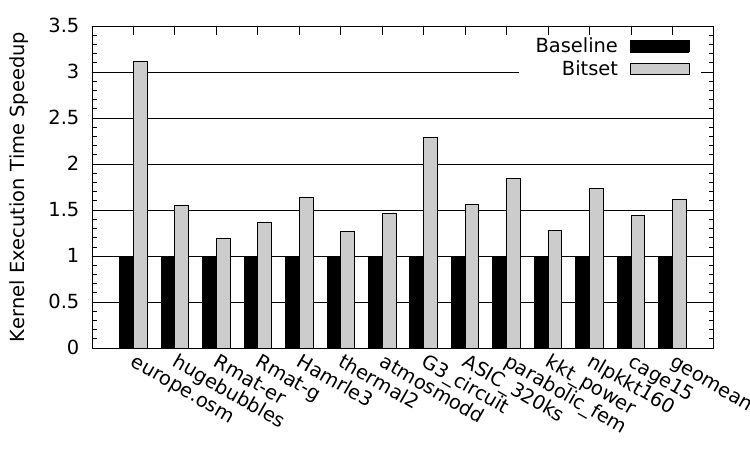}
		\caption{\texttt{FirstFit} kernel execution time speedup of \texttt{bitset}
		over the baseline. The kernel execution time is obtained by \texttt{nvprof}.}
		\label{fig:firstfit}
	\end{center}
\end{figure}

Fortunately, NVIDIA GPU architecture provides the \texttt{\_\_ffs()} intrinsic 
for our use. Find first set (ffs) or find first one is a bit operation that 
identifies the least significant index or position of the bit set to one in 
the word. So our scheme is to initialize the bits as all ``1''s, and clear the 
bit if the corresponding color is impermissible. To find the first permissible
color, we need only to call the \texttt{\_\_ffs()} intrinsic. This implemention
turns a for-loop into a single instruction and thus significantly reduces
the operations required to complete the \texttt{FirstFit} kernel. \cref{fig:firstfit} 
shows a 61\% speedup of the \texttt{FirstFit} kernel runtime over the baseline 
on average when \texttt{bitset} is applied. We also observe that this kernel 
improvement leads to an average of 28\% speedup of the entire program
compared to the baseline.

\subsection{Common Optimization Techniques}\label{subsect:cot}
Existing GPU graph processing algorithms have already utilized many optimization
techniques to improve performance. In graph coloring we employ some of these 
optimizations, including \textit{kernel fusion}, \textit{read-only data caching},
and \textit{load balancing} to enhance our implementation.

\begin{figure}[!t]
	\begin{center}
		\includegraphics[width=0.7\textwidth]{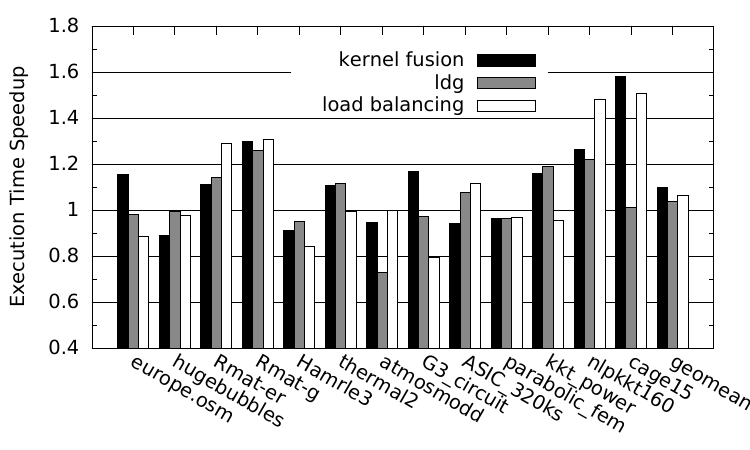}
		\caption{Program execution time speedup of \textit{kernel fusion}, 
			\textit{ldg} and \textit{load balancing} over the baseline.}
		\label{fig:fusion}
	\end{center}
\end{figure}

\textbf{Kernel Fusion.}
Previous techniques focus on individual kernels. However, another important
optimization technique called \textit{kernel fusion} combines multiple GPU
kernels into a single one, and thus can keep the entire program on the GPU.
Since adjacent kernels in CUDA share no state, this technique can leverage 
producer-consumer locality between operations and thus save significant 
memory bandwidth~\cite{Gunrock}. Note that global barrier is required between 
\texttt{FirstFit} and \texttt{ConflictResolve} operations. We use the existing 
method proposed by Xiao~\emph{et~al.}~\cite{Xiao}. With this global barrier,
kernels can only launch limited number of thread blocks, and thus thread 
coarsening is forced to be applied to both kernels. \cref{fig:fusion} shows an 
average 10\% speedup of kernel fusion over the baseline. As shown, benchmarks 
e.g. \texttt{cage15} and \texttt{rmat-g} can benefit from better locality brought 
by kerel fusion since they are relatively denser and more irregular than others. 
We observe an improved L2 cache hit rate for \texttt{cage15}.

\textbf{Read-only Data Caching.} In CUDA devices of compute capability 3.5 
and higher, data that is read-only for the entire lifetime of the kernel can 
be kept in the read-only data (unified L1/texture) cache by reading it using 
the intrinsic \texttt{\_\_ldg()}~\cite{CUDA}. We use the texture cache 
to hold the read-only data, i.e. the $C$ array and the $R$ array. And then more 
read-only data is forced to be cached in the L1 read-only cache whose access 
latency is around 30 cycles which is much shorter than the DRAM access latency 
(about 300 cycles). Therefore, \texttt{\_\_ldg()} can capture temporal locality 
and improve the performance because of reduced DRAM accesses. As shown in
\cref{fig:fusion}, \texttt{\_\_ldg()} can bring 3.6\% speedup over the baseline.

\textbf{Load Balancing.} 
Another important issue for graph algorithms is load imbalance.
The problem is particularly worse for scale-free (power-law) graphs.
Merrill~\emph{et~al.}~\cite{Merrill} proposed a hierarchical load
balancing strategy which maps the workload of a single vertex to 
a thread, a warp, or a thread block, according to the size of its 
neighbor list. At the fine-grained level, all the neighbor list 
offsets in the same thread block are loaded into shared memory, 
then the threads in the block cooperatively process per-edge 
operations iteratively. At the coarse-grained level, per-block
and per-warp schemes are utilized to handle the extreme cases: 
(1) neighbor lists larger than a thread block; (2) neighbor 
lists larger than a warp but smaller than a thread block 
respectively. We implement this strategy on graph coloring.
\cref{fig:fusion} illustrates the effect of load balancing on 
the benchmarks. Irregular benchmarks with uneven degree 
distribution, e.g. \texttt{rmat-g} and \texttt{cage15} can
substantially benefit from this technique. On average, it
achieves 6.4\% speedup over the baseline.

%% file: sections/sect-4.tex
\section{Evaluation}\label{sect:evaluation}

We use the R-MAT~\cite{R-MAT} graph generator to create synthetic 
graphs. The R-MAT algorithm determines the degree distribution by using 
four non-negative parameters (a; b; c; d) whose sum equals one. 
We generated two graphs (Rmat-er and Rmat-g) with 1M vertices 
size but varying structures by using the following set of parameters:
(0:25; 0:25; 0:25; 0:25); (0:45; 0:15; 0:15; 0:25). We also pick 
real-world sparse graphs from the University of Florida Sparse Matrix 
Collection~\cite{FSMC}. These benchmarks are also used in previous 
work~\cite{ILU,Merrill}. The matrices with the respective number 
of vertices (i.e. rows) and edges (non-zero elements) are shown in 
Table~\ref{table:bench}. The graphs vary widely in size, degree 
distribution, density of local subgraphs and application domain.

\subsection{Experiment Setup}
\begin{table*}[b!]
	\small
	\centering
	\begin{tabular}{| l | c | c | c | c | c |}
		\hline
		\bf{Name} & $\bm{n (10^{6})}$ & $\bm{m (10^{6})}$ & $\bm{\bar{d}}$ & $\bm{\sigma}$ & \bf{Description}\\
		\hline
		\texttt{europe.osm} & 50.9 & 108.1 & 2.1 & 0.23 & Road Network \\
		\hline
		\texttt{hugebubbles} & 21.2 & 63.6 & 3.0 & 0 & Adaptive Mesh\\
		\hline
		\texttt{rmat-er} & 1.0 & 10.0 & 10.0 & 10.83 & Synthetic\\
		\hline
		\texttt{rmat-g} & 1.0 & 10.0 & 10.0 & 123.34 & Synthetic\\
		\hline
		\texttt{Hamrle3} & 1.4 & 11.0 & 7.6 & 7.2 & Circuit Sim.\\
		\hline
		\texttt{thermal2} & 1.2 & 8.6 & 7.0 & 0.7 & Thermal Sim.\\
		\hline
		\texttt{atmosmodd} & 1.3 & 8.8 & 6.9 & 0.1 & Atmosphere\\
		\hline
		\texttt{G3\_circuit} & 1.6 & 7.7 & 4.8 & 0.4 & Circuit Sim.\\
		\hline
		\texttt{ASIC\_320ks} & 0.3 & 1.8 & 5.7 & 63.2 & Circuit Sim.\\
		\hline
		\texttt{parabolic\_fem} & 0.5 & 3.7 & 7.0 & 0.02 & General\\
		\hline
		\texttt{kkt\_power} & 2.1 & 14.6 & 7.1 & 54.8 & Optimization\\
		\hline
		\texttt{nlpkkt160} & 8.3 & 229.5 & 27.5 & 7.3 & Optimization\\
		\hline
		\texttt{cage15} & 5.2 & 99.2 & 19.2 & 32.9 & Electrophoresis \\
		\hline
	\end{tabular}
	\caption{Suite of benchmark graphs. \textnormal{$n$: number of vertices,
	$m$: number of edges, $\bar{d}$: average dergee, $\sigma$: degree variance.}}
	\label{table:bench}
\end{table*}

\begin{figure*}[!t]
	\begin{center}
		\includegraphics[width=\textwidth]{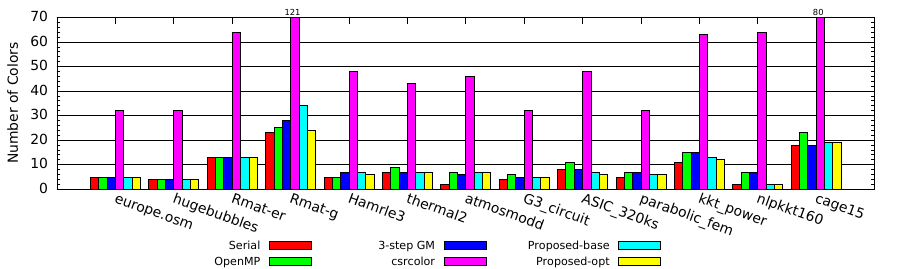}
		\caption{Total number of colors assigned with different implementations.}
		\label{fig:color}
	\end{center}
\end{figure*}

We compare 6 implementations including 
(1) \texttt{Serial}: the serial implementation in CUSP~\cite{CUSP}, 
(2) \texttt{OpenMP}: the baseline OpenMP implementation in~\cite{GCA},
(3) \texttt{3-step GM}: the previously proposed GM GPU implementation~\cite{EGC},
(4) \texttt{csrcolor}: the routine provided by NVIDIA CUSPARSE~\cite{ILU},
(5) \texttt{Proposed-base}: our proposed baseline data-driven implementation,
(6) \texttt{Proposed-opt}: our proposed optimized data-driven implementation.
We conduct the experiments on the NVIDIA K40c GPU with CUDA Toolkit 
7.5 release. \texttt{Serial} is executed on Intel Xeon E5-2690V2
2.30 GHz CPU with 12 cores. All the benchmarks are executed 10 
times and we collect the average execution time to avoid system 
noise. Timing is only performed on the computation part of each 
program. For all the GPU implementations, the input/output data 
transfer time (usually takes 10\%-20\% of the entire program execution time) 
is excluded because data is resident on the GPU in real applications~\cite{ILU}.

\subsection{Coloring Quality}
\cref{fig:color} shows the number of colors needed by different 
implementations for each graph. It is not surprising that implementations 
except \texttt{csrcolor} need similar amount of colors, since 
they are all based on the greedy scheme. The slight difference
among these 5 implementations may result from the different orderings
that are caused by different thread mapping stratigies and so on. 
\texttt{csrcolor}, however, needs 3.9$\times$$\sim$31$\times$
more colors than \texttt{Serial}, making this MIS based 
implementation unattractive or even unapplicable in many scenarios.
This substantial difference of coloring quality between \texttt{csrcolor} 
and the other implementations stems from the inherent algorithm
property of the SGR scheme and the MIS scheme. SGR uses greedy 
scheme, and for parallel versions it optimistically 
does coloring in parallel with later conflict resolve.
MIS, however, tries to find independent sets iteratively, 
which does not cause any conflict, but for performance concern, 
the methods used to find independent sets should be simple enough,
and thus generate solutions that are far away from the optimal.

\subsection{Performance}
\begin{figure*}[!t]
	\begin{center}
		\includegraphics[width=\textwidth]{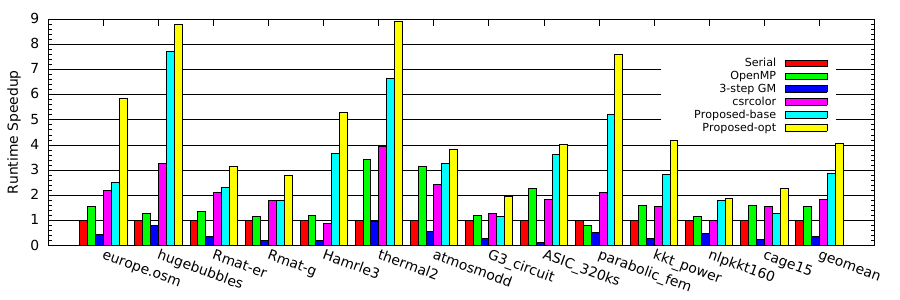}
		\caption{Runtime speedup normalized to the serial algorithm.}
		\label{fig:speedup}
	\end{center}
\end{figure*}

\cref{fig:speedup} illustrates the execution time speedup normalized to 
\texttt{Serial}. \texttt{OpenMP} on CPU achieves only moderate speedup 
(1.54$\times$). As mentioned before, \texttt{3-step GM} gets unacceptable 
performance: 62\% average slowdown compared to \texttt{Serial}. 
The slowdown stems from its mapping strategy and different data representation.
In contrast, \texttt{csrcolor} is a much faster GPU implementation. 
It achieves an average speedup of 1.84$\times$ over \texttt{Serial}. 
For regular graphs, such as \texttt{hugebubbles} and
\texttt{parabolic\_fem}, it performs much better than \texttt{OpenMP}. 
This shows the high throughput and bandwidth advantages of GPUs over CPUs.

Our proposed baseline implementation performs even better than 
\texttt{csrcolor}. We observe 2.87$\times$ speedup on average over 
\texttt{Serial}. It is 85.8\% and 56.1\% faster than \texttt{OpenMP} 
and \texttt{csrcolor} respectively. For some benchmarks e.g. 
\texttt{Hamrle3} and \texttt{parabolic\_fem}, \texttt{Proposed-base} 
significantly outperforms \texttt{csrcolor} (4.18$\times$ and 2.46$\times$). 
This performance boost mainly comes from the selection of data-driven 
algorithm structure and the atomic operation reduction. 
However, for relatively dense or irregular benchmarks, e.g. \texttt{cage15}, 
it performs worse than \texttt{csrcolor}, because no specific 
work is done to handle irregular cases and \texttt{csrcolor} 
has fewer memory accesses as mentioned before. 

With careful algorithm refinement and optimization techniques, 
we further improve the performance with an average speedup of 
4.08$\times$ over \texttt{Serial}. It is 2.63$\times$, 2.21$\times$ 
and 1.42$\times$ speedup over \texttt{OpenMP}, \texttt{csrcolor} 
and \texttt{Proposed-base} respectively. Generally, for regular 
benchmarks, it takes advantage of GPU's high throughput as 
\texttt{csrcolor} does, and performs even better because of the 
efficient bitset operator, fast convergence and so on, e.g.
\texttt{hugebubbles} (8.8$\times$) and \texttt{thermal2} (8.9$\times$).
For irregular benchmarks, better locality and load balance lead 
to better performance. Thus \texttt{Proposed-opt} can consistently 
outperform existing CPU and GPU parallel implementations.

We also notice that for some benchmarks, e.g. \texttt{G3\_circuit} 
and \texttt{nlpkkt160}, \texttt{Proposed-opt} gets very limited 
performance improvement compared to \texttt{Serial}. It is clear 
that the performance of graph coloring highly depends on the graph
characteristics (scale, density, degree distribution and topology).
For example, \texttt{nlpkkt160} has a relatively large average 
degree and suffers from conlicts. And some are small in size,
which limits the potential of performance improvement using GPUs.
But more importantly, since the compute operation is trivial, 
the performance is likely to be limited by memory operations.
For sparse graphs, not much temporal locality exists, and thus the 
kernel becomes extremely memory bound with large-scale datasets,
which could not be mitigated by the optimizations that 
we employ. We suggest system software or hardware support for 
efficient memory access to overcome this performance bottleneck.

\begin{figure}[!b]
	\begin{center}
		\includegraphics[width=0.7\textwidth]{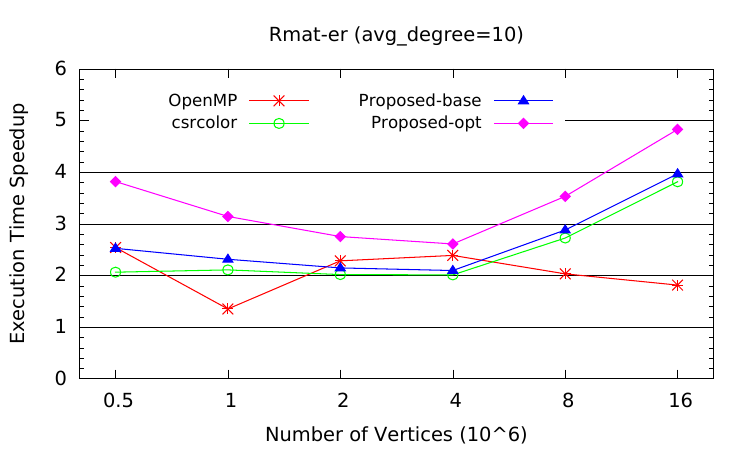}
		\caption{Execution time speedup of \texttt{Rmat-er} and \texttt{Rmat-g} 
			with various graph size (number of vertices), all normalized to \texttt{Serial}.}
		\label{fig:rmat-scale}
	\end{center}
\end{figure}

\subsection{Scalability}
To evaluate the scalability of our design on the input size, we vary the 
graph size (number of vertices) of \texttt{Rmat-er} from 500K to 16M with 
fixed average degree ($\bar{d}=10$). \cref{fig:rmat-scale} illustrates that 
\texttt{Proposed-opt} could achieve even more performance speedup given larger input datasets.
Our proposed implementation can consistently gain more than 2.5$\times$ 
speedup as the graph size changes, and always outperforms \texttt{csrcolor}.
After 4M vertices, the speedup increases siginificantly as the graph size 
increases, while \texttt{OpenMP} changes moderately and even drops at the
extremely large size. There is also a slight drop around 3M size for GPU 
impelementaions. This drop is related to the graph characteristics on 
which the performance highly depends on as mentioned. Here the cause is 
most likely the graph topology and degree distribution. Even so, 
\texttt{Proposed-opt} is still 9.3\% faster than \texttt{OpenMP} at
the 4M size, while it gets 2.66 speedup over \texttt{OpenMP} at the 16M 
size. For \texttt{Rmat-g} (not illustrated), we see a similar trend. 

\subsection{Sensitivity to Density} 
\begin{figure}[!t]
	\begin{center}
		\includegraphics[width=0.7\textwidth]{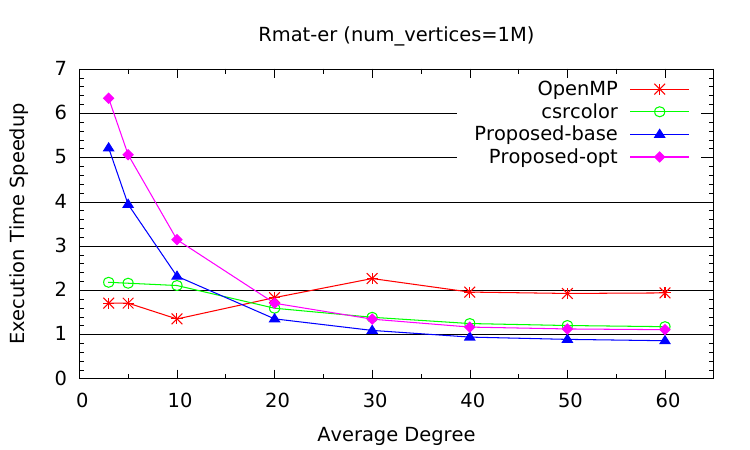}
		\caption{Execution time speedup of \texttt{Rmat-er} with various 
			average degrees, all normalized to \texttt{Serial}.}
		\label{fig:rmat-er-degree}
	\end{center}
\end{figure}

As mentioned, graph algorithms are highly sensitive to the characteristics
of the input datasets. We evaluate sensitivity to the graph density of our 
proposed graph coloring implementation. In \cref{fig:rmat-er-degree},
we vary the average degree $\bar{d}$ of \texttt{Rmat-er} with fixed 
graph size (1M vertices). We compare \texttt{OpenMP}, \texttt{csrcolor}, 
\texttt{Proposed-base} and \texttt{Proposed-opt}, all normalized to 
\texttt{Serial}. As shown, \texttt{Proposed-opt} significantly outperforms 
the others when $\bar{d}$ is small. This means our proposal can efficiently
handle sparse graphs. However, as the average degree increases, the performance
improvement over \texttt{Serial} decreases for \texttt{csrcolor} and our proposals. 
Their curves drop blow \texttt{OpenMP} when $\bar{d}$ is larger than 20. 

In contrast, \texttt{OpenMP} is more stable than GPU implementations.
The drop of GPU ones results from the conflicts between neighbors which
is not an issue in CPUs. As the graph becomes denser, the conflicts 
happen more frequently. In this case, the GPU implementations need much 
more iterations to complete than \texttt{OpenMP}. For dense graphs, 
thanks to the techniques that alleviate conflicts, \texttt{Proposed-opt}
still achieves comparable performance to \texttt{csrcolor}, while 
\texttt{Proposed-base} becomes worse than the other two GPU ones
and finally becomes slower than the serial implementation (blow 1). 
Remeber that our proposals still consistently yield much better coloring
quality than \texttt{csrcolor}. Although GPU implementations achieves 
high performance for sparse graphs, for dense graphs we suggest
to use CPUs instead of GPUs to solve the graph coloring problem.

%% file: sections/sect-5.tex
\section{Related Work}\label{sect:relatedwork}

Many graph algorithms have been developed on GPUs. 
Harish~\emph{et~al.}~\cite{Harish} are the pioneers to implement 
GPU graph algorithms. They developed topology-driven Breadth-first Search 
(BFS) and shortest path algorithms. Hong~\emph{et~al.}~\cite{Hong} proposed 
another topology-driven BFS to map warps rather than threads to vertices. 
Luo~\emph{et~al.}~\cite{Luo} developed the first work-efficient BFS on GPUs. 
Merrill~\emph{et~al.}~\cite{Merrill} improved Luo's work. They employed 
prefix sum to reduce atomic operations and used dynamic load balancing 
to deal with scale-free graphs. This implementation thus achieves high 
throughput and good scalability. The two major techniques 
of their work are also applicable to our implementation, while our work 
focuses more on the algorithm-specific refinement, e.g. the specific 
strategies to alleviate side effects of GPU's massive parallelism. 

Davidson~\emph{et~al.}~\cite{SSSP} developed a work-efficient
Single-Source Shortest Path (SSSP) algorithm on the GPU. They used another 
load balancing strategy which partitions the work into chunks and assigns 
each chunk to a block. Reserchers also proposed GPU implementations of Betweenness 
Centrality~\cite{BC}, Minimum Spanning Tree~\cite{Vineet,MST}, Strongly Connected 
Components~\cite{SCC} and so on. These work together demonstrated that with 
careful mapping and optimizations graph algorithms can get substantial 
performance boost on the GPU. Our work further enhances the conclusion of 
previous practices, while we show the importance of algorithm refinement
and architecture-specific optimizations for the problem of graph coloring.

Researchers have proposed many optimization techniques for graph algorithms,
or more generally, for irregular algorithms on GPUs. LAVER~\cite{LAVER} is 
a locality-aware vertex scheduling scheme which reorders the vertex queue 
to improve temporal locality of vertex data stored in on-chip caches.
Nasre~\cite{Atomic} proposed high-level methods to eliminate atomics in 
irregular programs, e.g. BFS and SSSP, on GPUs. Gunrock~\cite{Gunrock} 
absorbs previous knowledge and provides a library solution for GPU graph 
processing. It provides a load balancing framework based on Merrill's and 
Davidson's strategies, and integrates a set of common optimization techniques. 
A huge amount of efforts~\cite{Pregel,LIG,GraphLab,PowerGraph,Ligra,Medusa,Cusha,MapGraph}
have been made by researchers to generalize graph processing computation and reduce programmer's burden. 
Although generalized method can improve programmability, we argue that optimizations 
customized for the specific algorithm (which is difficult to generalize) is also important.

Che~\emph{et~al.}~\cite{Che} characterize a suite of GPU graph applications
and suggest architectural support. Xu~\emph{et~al.}~\cite{Xu} 
evaluate existing GPU graph algorithms on both a 
GPU simulator and a real GPU card and also suggest GPU hardware support. 
Wu~\emph{et~al.}~\cite{HLP} characterize three GPU graph frameworks 
and suggest to focus on constructing efficient operators.
Beamer~\emph{et~al.}~\cite{Beamer1} also measure three graph 
libraries and propose processor architecture change.
Green-Marl~\cite{Green-Marl} is a domain specific language for graph processing.
Chen~\emph{et~al.}~\cite{Chen:ERS} proposed compiler optimization methodology
for graph and other irregular applications on Intel Xeon Phi coprocessors.
Ahn~\emph{et~al.}~\cite{Ahn} developed a customized Processing-in-Memory (PIM) 
accelerator for large-scale graph processing. We believe that language, compiler, 
runtime and architecture support is necessary for large-scale graph processing.

%% file: sections/sect-6.tex
\section{Conclusion and Future Work}\label{sect:conclusion}
Graph coloring is an important graph algorithm that has been applied 
in many application domains. To process large-scale graphs, parallel graph 
coloring has been intensively studied in the past. Meanwhile, GPUs have 
been broadly utilized to speed up compute intensive kernels of
HPC applications in the past decade. In this paper, 
we explore parallel graph coloring on the GPU. 
Existing implementations either achieve limited performance or yield 
unsatisfatory coloring quality. We present a high performance graph 
coloring implementation for GPUs with good coloring quality. 
We utilize the speculative greedy scheme that guarantees coloring quality,
and improve performance with algorithm refinement and common 
optimization techniques. Experimental results show that our 
proposed implementation outperforms existing GPU implementations 
in terms of both performance and coloring quality. 
This work helps us further understand graph algorithms on modern massively
parallel processors and gives insight on the importance of both algorithm-specific 
and non-algorithm-specific (common) optimizations. We also show the
necessity of lower level support from system software and architecture.

In the future, we will further investigate the effect of conflict-resolution heuristics
on performance and coloring quality, and possibly propose even better heuristic.
We will also try to implement our proposal on Intel Xeon Phi coprocessors,
and try to optimize it for the MIC architecture. Besides, it would be interesting
to implement it on a GPU or MIC cluster to evaluate the scalability of our work.

\section*{Acknowledgment}
We thank the anonymous reviewers for their insightful comments and suggestions,
and A.V. Pascal Grosset from University of Utah for generously sharing his source code.
This work is partly supported by the National Natural Science Foundation of China 
(NSFC) under grant No.61502514, No.61602501, No.61402488, and No.61502509, 
and the National Key Research and Development Program of China under grant No.2016YFB0200400.